\documentclass[fleqn,10pt]{wlscirep}
\usepackage[utf8]{inputenc}
\usepackage[T1]{fontenc}

\usepackage{morefloats}
\usepackage{mathrsfs} 
\usepackage{textcomp}
\usepackage{amsmath}
\usepackage{braket}

\newcommand{\bq}{\boldsymbol{{\rm q}}}
\newcommand{\bA}{\boldsymbol{{\rm A}}}
\newcommand{\br}{\boldsymbol{{\rm r}}}

\title{Quantized Piezospintronic Effect in Moir\'e Systems}

\author[1]{Mario Castro}
\author[1]{Benjamín Mancilla}
\author[1]{ Fabian Wolff}
\author[1,*]{Alvaro S. Nunez}

\affil[1]{Departamento de F\'isica, FCFM, Universidad de Chile, Santiago, Chile.}

\affil[*]{alnunez@dfi.uchile.cl}

\begin{abstract}
This paper presents a novel approach for generating and controlling spin currents in an antiferromagnetic twisted honeycomb bilayer in response to an elastic deformation. Utilizing a continuum model, closely based upon the seminal Bistritzer-MacDonald model, that captures the essential physics of low-energy moiré bands, we calculate the spin current response to the deformation in terms of the familiar Berry phase formalism. The resulting moiré superlattice potential modulates the electronic band structure, leading to emergent topological phases and novel transport properties such as quantized piezo responses both for spin and charge transport. This approach allows us to tune the system across different topological regimes and to explore the piezo-spintronic responses as a function of the band topology. When inversion symmetry is broken either by a sublattice potential $V$, alignment with an hBN substrate, uniaxial strain, or structural asymmetry present in the moiré superlattice, the system acquires a finite Berry curvature that is opposite in the $K$ and $K'$ valleys (protected by valley time reversal symmetry). In contrast, for strain, the valley-contrasting nature of the pseudo-gauge field ensures that the quantized response is robust and proportional to the sum of the valley Chern numbers. These notable physical properties make these systems promising candidates for groundbreaking spintronic and valleytronic devices.
\end{abstract}
\begin{document}

\flushbottom
\maketitle
%
%
\thispagestyle{empty}

\section*{Introduction}

The capacity to generate and exert control over spin currents\cite{Maekawa2017} is a fundamental aspect of spintronics\cite{Xu2015, Dey2022}, a field that has witnessed steady progress during the last decades by aiming to transform electronics by harnessing the intrinsic angular momentum of electrons—referred to as spin—alongside their charge. Traditional electronics, which depended solely on charge transport, encountered limitations regarding power consumption and the potential for miniaturization. In contrast, spin currents present an attractive solution for engineering devices that are not only faster and more energy-efficient but also possess non-volatile properties. As a result, the creation of stable and effective spin current sources is of strategic importance for advancing future information technologies. The endeavor to develop varied and manageable sources of spin currents, whether achieved through electrical means, thermal gradients—acknowledged as the spin Seebeck effect—mechanical deformations known as the piezospintronic effect, or excitation via optical methods, is critically significant. Each innovative technique for generating and controlling spin currents enhances the arsenal available to spintronics engineers and physicists, progressively moving us toward a novel era characterized by high-performance electronic devices that operate with minimal power consumption.

A phenomenon has been proposed where certain materials can develop a pure spin current in response to mechanical strain\cite{Nunez2014}. This effect, termed piezospintronics, is analogous to the well-known piezoelectric effect\cite{Curie1880, Gautschi2002, Cady2018}, where materials become electrified when subjected to strain. The authors provided the theoretical framework for this effect in subsequent articles, where a discussion of the necessary symmetry requirements was presented, along with an illustration of the concept using several model systems \cite{Nunez2014,Ulloa2017,Castro2024,Vergara2024}.

The piezo-spintronic effect is predicted to occur in materials that lack inversion symmetry. Unlike charge currents, which are odd under time reversal, the spin current is even. This effect also requires the breaking of time reversal symmetry. The theoretical response can be represented geometrically using spin Berry phases\cite{Nunez2014}, drawing a close analogy with the theory of electric polarization and the piezoelectric effect\cite{Vanderbilt2018}. The effect can be readily cast using the specific definition of spin current, 
$J^S_{ij}={\rm d}P^S_{ij}/{\rm d}t$. This definition is linked to the time derivative of the spin dipolar moment and is useful because it reduces to the intuitive notion of spin current when spin is conserved.

The piezo-spintronic effect can be understood as two separate, opposite piezoelectric effects, one for each spin channel. Under strain, this leads to opposite currents for each spin, resulting in a net spin current but no net charge current. This phenomenon is predicted to occur in crystals that are invariant under the consecutive action of both spin reversal (R) and spatial inversion (I) operators. The theory suggests that the presence of spin-orbit interaction is not necessary for this effect to be displayed.
Antiferromagnetic graphene satisfies those symmetry requirements and was one of the first models where the effect was predicted\cite{Ulloa2017}.
A second model fulfilling the symmetry requirements is a spin-dependent generalization of the Avron-Berger-Last (ABL)\cite{Avron1997} model for quantum piezoelectricity. This model involves a triangular lattice with different magnetic fluxes piercing up and down triangles. The hopping amplitudes are linked to elastic deformations, and the model can be shown to display a topologically quantized piezo-spintronic response\cite{Nunez2014}. In complementary papers, the models have also served as a basis for toy representations of multiferroic systems\cite{Saez2023, Saez2024, Castro2024}. Lately, a family of materials that might display the effect was proposed\cite{Chen2025}.

Magnetic moiré\cite{Yao2024, Ciorciaro2023, AkifKeskiner2024, Liu2024, Slot2023} and multiferroic moiré\cite{Fumega2023, Abouelkomsan2024, Sun2022, Sun2023} systems represent a burgeoning field with the potential to revolutionize spintronics by offering new ways to control and manipulate spin through the intricate interplay of moiré patterns and their collective order. The ongoing research in this area promises exciting discoveries and the development of innovative spintronic technologies. For this reason, it is interesting to pursue the notion of piezospintronics moiré systems. This is the main content of the present letter. We will argue that, due to the topological peculiarities of the physics of twisted bilayers, as encoded in the seminal Bistritzer-MacDonald formalism \cite{Bistritzer2011}, the piezospintronic coefficient of the system turns out to be quantized. As we will discuss, this opens great opportunities for the generation of spin currents and, conversely, elevating precise measurements of the spin currents toward unprecedented degrees. These notable physical properties make these systems promising candidates for groundbreaking spintronic and valleytronic devices.

\section*{Results and Discussion}
\subsection*{Continuum Model}
To describe the interplay between topology, strain, and spin-dependent effects in an antiferromagnetic twisted honeycomb bilayer\cite{Andrei2020, LopesdosSantos2012, LopesdosSantos2007, He2013}, we employ a continuum model that captures the essential physics of low-energy moiré bands\cite{Bistritzer2011, Watson2023, Becker2024}. This model is constructed by combining Dirac Hamiltonians for each layer, incorporating a relative twist angle and a uniaxial strain applied to the bottom layer, defined as $\mathcal{E} = R_{-\phi} \begin{pmatrix}
-\epsilon& 0 \\
0 & \nu \epsilon\\
\end{pmatrix}R_{\phi}$ with $\nu$ is the Poisson ratio and $\phi$ is the angle of the uniaxial strain relative to the zig-zag direction. The resulting moiré superlattice potential significantly modifies the electronic band structure, leading to emergent topological phases and novel transport properties such as quantized (anomalous) Hall conductivity and spin-resolved charge transport, among others \cite{Serlin2020, Xiao2020}. In our approach, the interlayer coupling is described by the Bistritzer-MacDonald formalism \cite{Bistritzer2011}, while the strain is introduced as a strain tensor acting on the bottom layer, which enters the Hamiltonian as an effective gauge field with opposite sign in each valley. Additionally, we include a sublattice site potential $V$, which can be induced, for example, by alignment with a h-BN substrate, see Methods for more details. The presence of $V$ breaks the $C_{2z}$ symmetry (sublattice exchange or inversion through the plane) and reduces the point group symmetry from $D_6$ to $D_3$ \cite{He2020, Zhang2022, Arora2021}. In addition, it opens a gap between the valence and conduction bands and leads to a nontrivial Berry curvature. These ingredients allow us to tune the system across different topological regimes and to explore the piezoelectric and piezospintronic responses as a function of the band topology. 

\begin{figure}[!h]
\centering
\includegraphics[width=1.0\linewidth]{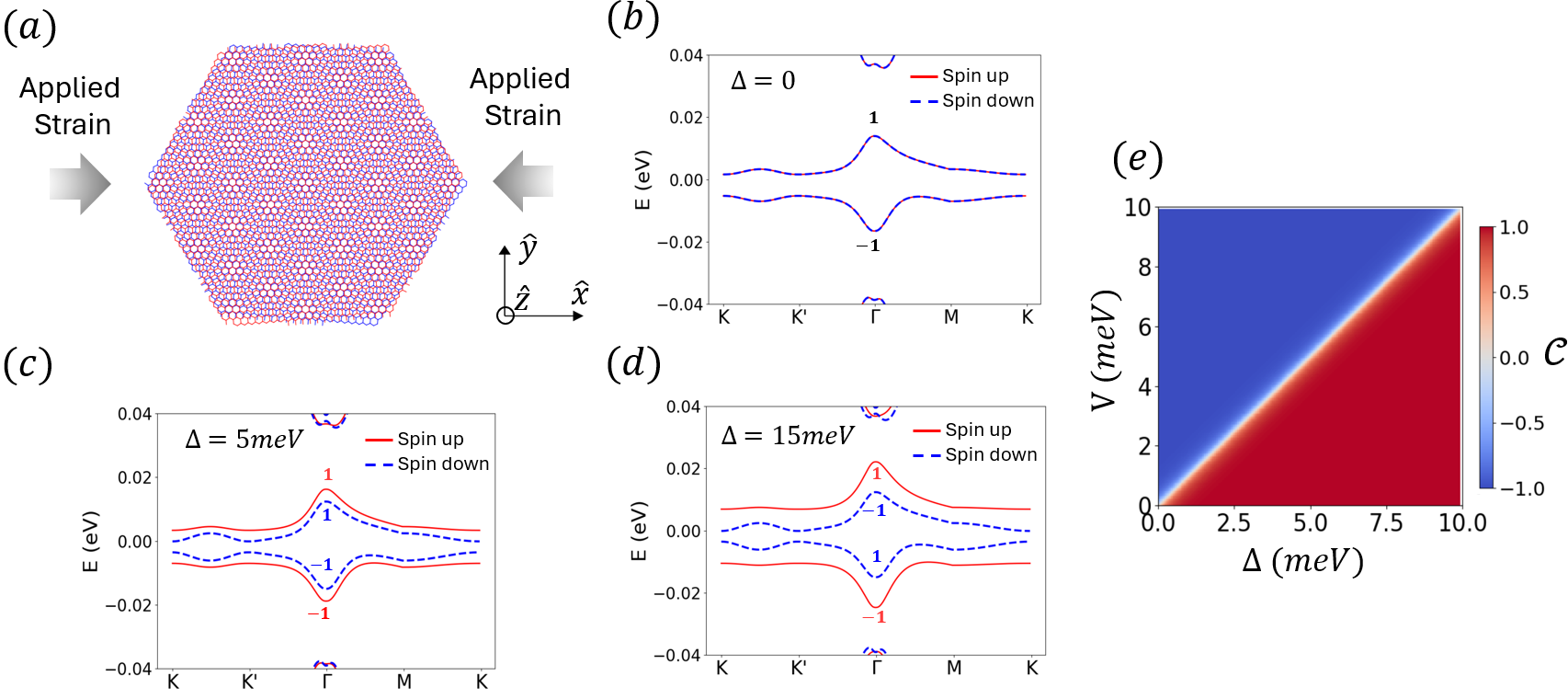}
\caption{ (a) Twisted honeycomb bilayer under uniaxial strain. (b-d) shows the band structure of the system for different values of $\Delta$. (b), (c) and (d) considers $\Delta = 0$, $\Delta = 5$ meV, and $\Delta = 15$ meV respectively. In addition, we consider $V = 10$ meV and $\xi = 1$. The red and blue colors denote the bands associated with spin up and down, respectively. (e) shows the Chern number of the spin-down valence band as a function of $V$ and $\Delta$. Depending on the sign of $V-\Delta$, the valley Chern number of the band can be $+1$ or $-1$.}
\label{fig:fig1}
\end{figure}

A schematic illustration of the twisted bilayer geometry and its low-energy band structure is shown in Fig. \ref{fig:fig1}. Figure \ref{fig:fig1}.a shows the top and bottom honeycomb lattices are depicted with a relative twist, and uniaxial strain is applied to the bottom layer. Figures \ref{fig:fig1}.(b–d) display the calculated band structure for $V = 10$ meV at $\Delta = 0$, $5$, and $15$ meV. In the absence of exchange-sd interaction ($\Delta = 0$), the sublattice potential $V$ lifts the sublattice degeneracy but preserves spin degeneracy. When $\Delta$ is finite, the spin degeneracy is lifted, resulting in well-resolved spin-up (red) and spin-down (blue) bands. This shows how the interplay between $V$ and $\Delta$ leads to spin-split moiré minibands, as captured by the model Hamiltonian. The topological properties of these minibands can be characterized by the valley Chern number $\mathcal{C}$, which is defined as $    \mathcal{C}_{n,\tau}^{\xi}= \dfrac{1}{\pi } {\rm Im}\left[\int_{mBZ}d^2\mathbf{k}\left\langle \frac{d\psi_{n\tau}^{\xi}}{dk_x} \,\middle|\, \frac{d\psi_{n\tau}^{\xi}}{dk_y} \right\rangle
\right]$, where $\ket{\psi_{n,\tau}^{\xi}}$ is the Bloch eigenstate of band $n$, spin $\tau$, valley $\xi$, and the integration is over the moiré Brillouin zone (mBZ). Moreover, the topological character of the moiré minibands depends sensitively on the relative strength of $V$ and $\Delta$. For $|V| > |\Delta|$, the two lower (upper) bands, regardless of spin orientation, carry the same valley Chern number $\mathcal{C} = -1$ ($+1$). In contrast, when $|\Delta| > |V|$, the Chern numbers of the spin down band change the sign, so that the two lower (or upper) bands have opposite valley Chern numbers. This behaviour can be qualitatively understood by examining the Berry curvature that can be approximated by obtaining the effective Hamiltonian for the low energy regime \cite{Cao2020}:

\begin{align}
    \Omega_{s}(q, \xi) = -\xi \dfrac{m_s \hbar^2 v_f^2}{2 (q^2 \hbar^2 v_f^2 + m_s^2)^{3/2}},
    \label{eq:Berry_curvature}
\end{align}

where $m_{\uparrow} = V +\Delta$ and  $m_{\downarrow} = V  - \Delta$ (for simplicity, we have not considered the strain effects and $u = u'$). For the spin-down case, the sign of the Berry curvature depends directly on the relative magnitudes of $V$ and $\Delta$ (see figure \ref{fig:fig1}.(e)). The resulting band inversion and associated topological transitions play a crucial role in determining whether the quantized response is piezoelectric or piezospintronic, as will be discussed in detail below.


\subsection*{Quantized piezoelectric and piezospintronic responses}

How a crystal reacts to mechanical stress or deformation can be effectively characterized by observing both the electric and spin dipole moments that are induced as a result. The concept of electric polarization, denoted by $\mathbf{P}^{e}$, specifically refers to the electric dipole moment within each unit cell of the crystal. Within the framework of modern polarization theory, this is conceptualized through the Berry phase associated with the electronic wavefunctions \cite{Yu2020, KingSmith1993}. In a similar vein, the spin dipole moment, represented as $\mathbf{P}^s$, measures the cumulative spin moment present within a unit cell \cite{Nunez2014}. This, too, can be described using the formalism of a spin Berry phase, aligning with its electrical analog.
\begin{align}
\gamma_{ijk}^{e} = \left. \frac{\partial \mathbf{P}^e_i}{\partial \epsilon_{jk}} \right|_{\epsilon=0} &&
\gamma_{ijk}^{z,s} = \left. \frac{\partial \mathbf{P}^{z,s}_i}{\partial \epsilon_{jk}} \right|_{\epsilon=0},
\end{align}

where $P^e_i$ and $P^{z,s}_{i}$ are the components of the electric and spin dipole moments, respectively, and $\epsilon_{jk}$ denotes the strain tensor components. Within the modern theory, these response tensors can be written in terms of Berry phase derivatives over the moiré Brillouin zone:

\begin{align}
    &\gamma_{ijk}^{e}= \dfrac{e}{2 \pi^2 }\sum_{n, \xi, \tau}{\rm Im}\left[\int_{mBZ}d^2\mathbf{k}   \left\langle \frac{d\psi_{n\tau}^{\xi}}{dk_i} \,\middle|\, \frac{d\psi_{n\tau}^{\xi}}{d\epsilon_{jk}} \right\rangle
\right],\\
    &    \gamma_{ijk}^{z,s} = -\dfrac{\hbar}{ (2 \pi)^2 } \sum_{\substack{n, \xi \\ \tau, \tau'}} {\rm Im}\left[\int_{mBZ}d^2 \mathbf{k}\bra{\dfrac{d\psi_{n\tau}^{\xi}}{dk_i}}\dfrac{ \sigma_{\tau\tau'}^{z}}{2}\ket{\dfrac{d\psi_{n\tau'}^{\xi}}{d\epsilon_{jk}}}\right],
\end{align}

where the sum in $n$ is performed over the occupied band. A nonzero piezoelectric response requires broken inversion symmetry, which in our system is provided by the staggered potential $V$. In contrast, the piezospintronic response requires simultaneous breaking of both inversion and time-reversal symmetry, achieved here by the combined action of $V$ and the antiferromagnetic exchange s-d interaction $\Delta$ \cite{Nunez2014,Ulloa2017}.   Importantly, in moiré systems with valley-contrasting gauge fields, these responses can become quantized. It is worth mentioning that while $\Delta$ breaks time-reversal symmetry in the spin sector, the valley symmetry is preserved.  The underlying $D_3$ point group symmetry of the moiré lattice further restricts the tensor structure. In what follows, we focus on the $\gamma_{yxx}$ component, which captures the quantized response relevant for our system \cite{Ulloa2017}. When the Fermi level lies within a topological gap, the piezoelectric \cite{Peng2022} and piezospintronic coefficients take quantized values determined by the sum of (valley and spin) Chern numbers of the two valence bands  (spin up and down contributions, see Fig \ref{fig:fig1})

\begin{align}
    \gamma_{yxx}^{e}\approx - (\zeta _{ijk}^{e}C_{\uparrow} +  \zeta _{ijk}^{e}C_{\downarrow}) &&     \gamma_{yxx}^{z,s}\approx  -(\zeta _{ijk}^{z,s}C_{\uparrow} -  \zeta _{ijk}^{z,s}C_{\downarrow})
    \label{quantization}
\end{align}

where $\zeta _{yxx}^{e} \approx \dfrac{e}{\pi} \dfrac{\sqrt{3}\beta}{a}$ and $\zeta _{yxx}^{z,s}\approx \dfrac{\hbar}{4\pi} \dfrac{\sqrt{3}\beta}{a}$ are parameters that depend on the properties of the monolayer. \\

\begin{figure}
    \centering
    \includegraphics[width=1.0\linewidth]{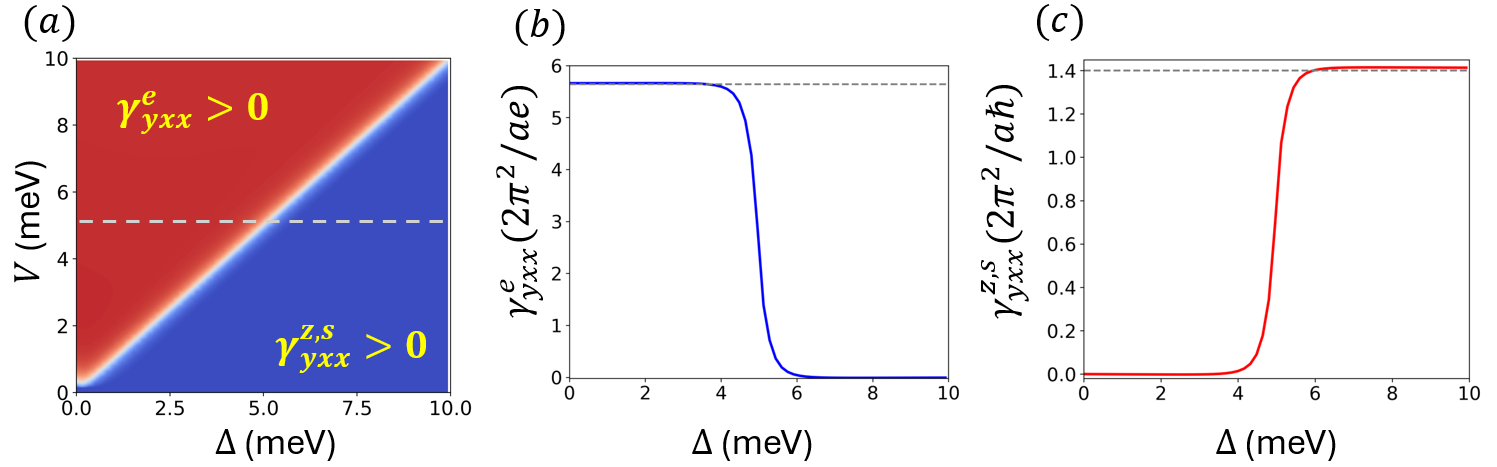}
    \caption{Piezoelectric and piezospintronics response as a function of $\Delta$. The gray dot-line denotes the value of the ideal quantized value \ref{quantization}. We have considered that V = 10 meV and $\xi = 1$. }
    \label{fig:fig2}
\end{figure}

In Fig. \ref{fig:fig2}, we present the calculated piezoelectric and piezospintronic responses as a function of the exchange-sd parameter. When $V > \Delta$, the Chern numbers of the two valence bands have the same sign, resulting in a quantized piezoelectric response and a suppressed piezospintronic response. Conversely, when $\Delta > V$, the Chern numbers alternate in sign between the two valence bands, leading to a quantized piezospintronic response while the piezoelectric response vanishes. It is important to note that, according to Eq. \ref{eq:Berry_curvature}, the Berry curvature has opposite sign in the $K$ and $K'$ valleys. However, the strain-induced pseudo-gauge field also couples with opposite sign to each valley, thus compensating the sign reversal of the Berry curvature. As a result, the contributions from both valleys add constructively, yielding a robust quantized piezoelectric (or piezospintronic) response, depending on the topological regime.  This mechanism is analogous to the quantization observed in the Hall effect, but with an important distinction in the Hall response, the electromagnetic field couples identically to both valleys, leading to a cancellation of their Berry curvature contributions \cite{Yu2020, Peng2022}. In contrast, for strain, the valley-contrasting nature of the pseudo-gauge field ensures that the quantized response is robust, and it can be approximated by the sum of the valley Chern numbers.  The value of the expressions \ref{quantization} is shown in a dot-gray line in the figure \ref{fig:fig2}(b-c). We observe a good agreement with the continuum model.


\subsection*{Orbital magnetism }

\begin{figure}
    \centering
    \includegraphics[width=0.7\linewidth]{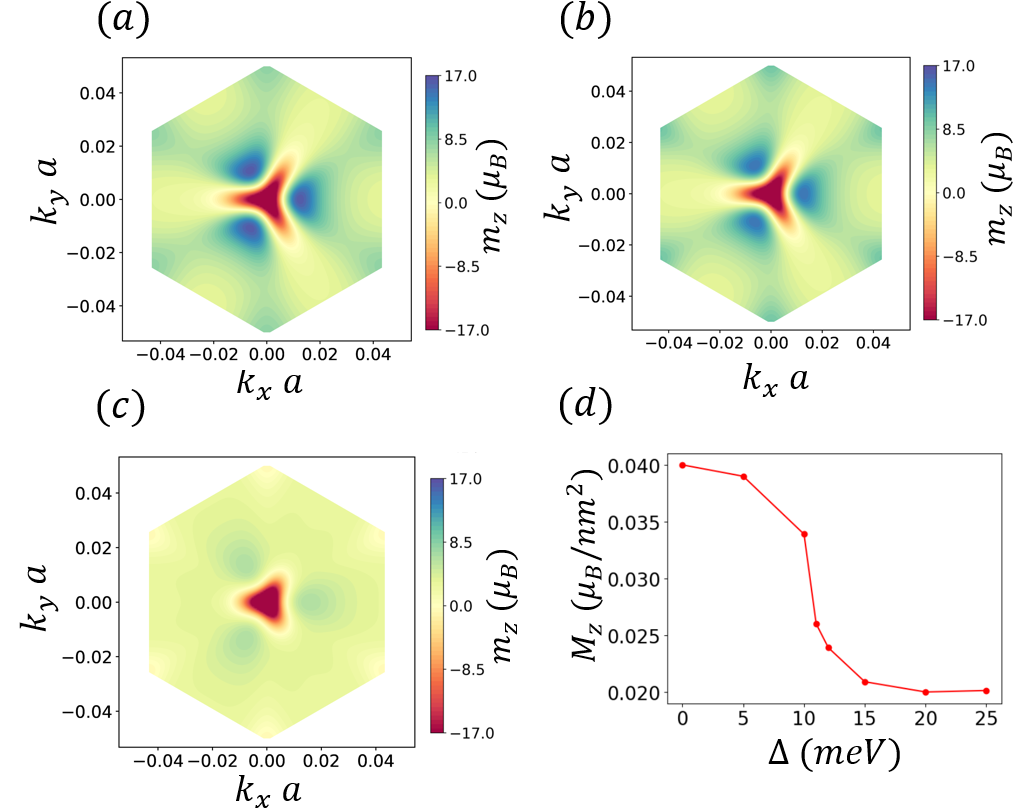}
    \caption{Orbital magnetic moment for differents value of $\Delta$ for the valence band.
    (a), (b) and (c) considers $\Delta = 0$, $\Delta = 5$ meV and $\Delta = 15 $ meV, respectively. We consider that $V = 10$ meV and $\xi = 1$. The color scale indicates the magnitude of the orbital moment (in units of $\mu_B$). (D) shows the total magnetizacion $M_z$ as a function of $\Delta$ showing a clear decrease of $M_z$ as $\Delta$ increases. }
    \label{fig:fig3}
\end{figure}

Recent experimental and theoretical works have established that moiré superlattices, such as twisted bilayer graphene (TBG) aligned with hBN, rhombohedral trilayer graphene, and twisted  bilayer graphene, can realize robust forms of orbital ferromagnetism, even in the absence of spin–orbit coupling \cite{Jadaun2023, Sharpe2019, Tschirhart2021}. In these systems, the observed magnetism is predominantly orbital in nature, leading to the notion of orbital Chern insulators and a range of topological phenomena associated with the Berry curvature of the electronic bands.  In our model, breaking inversion symmetry, via a sublattice potential $V$, alignment with an hBN substrate, uniaxial strain, or intrinsic structural asymmetry of the moiré lattice, leads to a finite Berry curvature with opposite sign in the $K$ and $K'$ valleys. This valley-contrasting Berry curvature is protected by valley time-reversal symmetry and manifests as orbital magnetization of equal magnitude but opposite sign in each valley, i.e., $m_{n\tau,+1}^z(\mathbf{q}) = -m_{n\tau,-1}^z(-\mathbf{q})$. The orbital magnetic moment per valley is given by

\begin{align}
    m_{n\xi}^{z, orb}(\mathbf{q}) = \frac{i e}{2 \hbar} \sum_{\tau}\bra{\dfrac{\partial u_{n\tau}^{\xi}(\mathbf{q})}{\partial \mathbf{q}} }\times (\mathcal{H}_{\xi}(\theta) - \mathcal{E}_{n \xi })\ket{\dfrac{\partial u_{n\tau}^{\xi}(\mathbf{q})}{\partial \mathbf{q}} },
\end{align}

where $\mathcal{E}_{n \xi }$ are the eigenvalue at zero field. The total magnetic moment combines both orbital and spin contributions:

\begin{align}
    m_{n\xi}^{z, total}(\mathbf{q}) = m_{n\xi}^{z, orb}(\mathbf{q}) + m_{n \xi}^{z, spin}(\mathbf{q}),
\end{align}

where the spin moment reads $m_{n\xi}^{z,\mathrm{spin}}(\mathbf{q}) = \langle u_{n\tau}^\xi(\mathbf{q}) | \frac{\hbar}{2} \sigma^z | u_{n\tau}^\xi(\mathbf{q}) \rangle$. At half-filling, spin-up and spin-down states are equally occupied, so the net spin magnetization per valley vanishes, and the total valley magnetization is determined by the orbital contribution  $M_{z,\xi} =  \sum_n \int  \dfrac{d^2\mathbf{q}}{(2\pi)^2}  m_{n\xi}^{z, total}(\mathbf{q})$.  To elucidate the interplay between antiferromagnetic exchange and valley orbital magnetization, Fig.~\ref{fig:fig3} presents the calculated distribution of the orbital magnetic moment $m_z(\mathbf{q})$ across the moiré Brillouin zone for different values of $\Delta$. The maps exhibit a threefold rotational ($C_{3z}$) symmetry, characteristic of the moiré superlattice in the absence of strain. For $\Delta = 0$, the orbital moment is strongly concentrated around the moiré $\Gamma$ point, with a pronounced $C_{3z}$-symmetric pattern. As $\Delta$ increases, the overall magnitude of $m_z(\mathbf{q})$ diminishes throughout much of the mBZ, as reflected in a reduction of the total orbital magnetization $M_z$. Importantly, the orbital magnetic moments remain particularly large around the $\Gamma$ point even for finite $\Delta$, consistent with previous theoretical reports that the largest orbital moments in moiré systems are localized near high-symmetry points, where the flat bands hybridize with adjacent bands \cite{He2020, Guerci2021}. This persistence highlights the interplay between band topology, moiré symmetry, and antiferromagnetic exchange, and points to the robust valley-contrasting magnetization in these systems.  The pronounced stability of the orbital magnetic texture, particularly around the moiré $\Gamma$ point, is protected by valley time-reversal symmetry inherent to the antiferromagnetic exchange configuration. Breaking this protection through targeted perturbations such as non-uniform strain fields that generate valley-dependent pseudomagnetic fields, valley-selective scattering channels, or in-plane electric currents can create conditions under which the otherwise compensated valley contributions give rise to a measurable macroscopic orbital magnetization \cite{Chen2023, He2020, Farajollahpour2017} . These scenarios provide a solid foundation for future experimental and theoretical efforts aimed at actively controlling orbital magnetism in moiré materials.

\section*{Conclusions}

This paper presents a novel approach for generating and controlling spin currents in an antiferromagnetic twisted honeycomb bilayer in response to an elastic deformation. The interlayer coupling is described by the seminal Bistritzer-MacDonald formalism, while the strain is introduced as a strain tensor acting on the bottom layer, which enters the Hamiltonian as an effective gauge field with opposite sign in each valley. The resulting moiré superlattice potential modulates the electronic band structure, leading to emergent topological phases and novel transport properties such as quantized piezoresponses for both spin and charge resolved transport, among others. In contrast, for strain, the valley-contrasting nature of the pseudo-gauge field ensures that the quantized response is robust and proportional to the sum of the valley Chern numbers. These significant physical characteristics render these systems as strong candidates for innovative advancements in spintronic and valleytronic technology. The pronounced stability of the orbital magnetic texture, particularly around the moiré $\Gamma$ point underscores its potential as a platform for engineering finite net orbital magnetization. In light of these findings, suitable experimental platforms for realizing the proposed antiferromagnetic twisted honeycomb bilayers include van der Waals intrinsic magnetic semiconductors from the MPX$_3$ family (M = Fe, Mn, Co, Ni; X = S, Se) \cite{Jing2023}, such as MnPS$_3$ and MnPSe$_3$ \cite{Li2013}, which host Néel-type order on a honeycomb lattice and are available down to the monolayer limit\cite{Rahman2021, Calder2021}. These layers can be incorporated into heterostructures where the second honeycomb layer is composed of an electronically inert or wide-gap material, such as hexagonal boron nitride (hBN) or SiC, providing a large staggered sublattice potential through their ionic character\cite{Lee2020}.

\section*{Methods}

We model the low-energy electronic structure of the antiferromagnetic twisted honeycomb bilayer using a generalized continuum Hamiltonian \cite{Bistritzer2011, Koshino2018} that includes twist, uniaxial strain, and exchange-sd inrteraction. Each layer is described by a Dirac Hamiltonian rotated by $\pm\theta/2$, with the bottom layer additionally subject to a uniaxial strain tensor $\mathcal{E} = R_{-\phi} \begin{pmatrix}
-\epsilon& 0 \\
0 & \nu \epsilon\\
\end{pmatrix}R_{\phi}$ with $\nu$ is the Poisson ratio and $\phi$ is the angle of the uniaxial strain relative to the zig-zag direction. In the hamiltonian, strain enters as a valley-contrasting gauge field $\mathbf{A}$, while a staggered sublattice potential $V$ (e.g., from hBN alignment) breaks $C_{2z}$ symmetry, opening a topological gap between the middle plane bands.  Aditionally, we consider an antiferromagnetic exchange-sd with strenght $\Delta$ acts in the spin sector. The low-energy Hamiltonian for valley $\xi = \pm 1$ and spin $\tau = \uparrow,\downarrow$ is:

\begin{align}
    \mathcal{H}_{\xi}(\theta) =\begin{pmatrix}
H_{b,\uparrow}^{\xi} (-\theta/2) & U & 0& 0\\
U^{\dagger} & H_{t,\uparrow}^{\xi}(\theta/2)& 0& 0\\
0& 0 &H_{b,\downarrow}^{\xi}(-\theta/2) & U \\
0& 0 & U^{\dagger} & H_{t,\downarrow}^{\xi}(\theta/2)\\
\end{pmatrix}
\end{align}
where $H_{b,\tau}^{\xi}(\theta) $ and $H_{t,\tau}^{\xi}(\theta) $ denote the Dirac-like hamiltonians for the bottom (strained) and top (unstrained) layers rotated in $\theta$, respectively. Additionally, $U$  corresponds to the moiré potential.

\begin{align}
    \mathcal{H}_b &= \sum_{\bq,\xi,\tau} a^{\dagger}_{b,\tau,\xi}(\bq)
    \left[
        \hbar v_f \mathcal{R}_{-\theta/2}(1+\mathcal{E}^{T}) (\bq + \xi \bA )\cdot \boldsymbol{\sigma} + V \sigma^{z}
    \right]
     \nonumber \\
    &\quad a_{b,\tau,\xi}(\bq)+ \Delta \eta_{\alpha} \sum_{\bq,\xi,\tau,\tau'} a^{\dagger}_{b,\tau,\xi} \sigma^{z}_{\tau,\tau'} a_{b,\tau',\xi} \nonumber \\
    \mathcal{H}_t &= \sum_{\bq,\xi,\tau} a^{\dagger}_{t,\tau,\xi}(\bq)
    \left[
        \hbar v_f \mathcal{R}_{\theta/2} \bq \cdot \boldsymbol{\sigma} + V \sigma^{z}
    \right]
    a_{t,\tau,\xi}(\bq) \nonumber \\
    &\quad + \Delta \eta_{\alpha} \sum_{\bq,\xi,\tau,\tau'} a^{\dagger}_{t,\tau,\xi} \sigma^{z}_{\tau,\tau'} a_{t,\tau',\xi}\\
    U &= \begin{pmatrix}
 u& u' \\
u' & u\\
\end{pmatrix}e^{-i \xi \bq_1 \cdot \br} + \begin{pmatrix}
 u & u' e^{-i\xi \lambda}\\
u'e^{i\xi \lambda} & u\\
\end{pmatrix}e^{-i \xi \bq_2 \cdot \br} \nonumber\\
&\quad+ \begin{pmatrix}
 u& u'e^{i\xi \lambda} \\
u'e^{-i\xi \lambda} & u\\
\end{pmatrix}e^{-i \xi \bq_3 \cdot \br},
\end{align}

with $\boldsymbol{\sigma}$ Pauli matrices, $\bq = \mathbf{k} - (1-\mathcal{E}^{T})K_{\xi}$  is the momentum relative to the Brillouin zone corner $K_{\xi}$,  and $\eta = \pm 1$ for each sublattice. In addition, $\lambda = 2\pi/3$ is the moiré superlattice period, $\bA = \dfrac{\sqrt{3} \beta}{a} (\epsilon_{xx} - \epsilon_{yy}, -2\epsilon_{xy})$ is the effective induced strained field and $\epsilon_{ij}$ the component of the strain tensor. Additionally, the hopping vector $\mathbf{q}_j$ are defined as:

\begin{align}
    &q_1 = -\dfrac{4\pi}{3\sqrt{3} d} (\epsilon_{xx}\cos(\theta/2), (2-\epsilon_{xx})\sin(\theta/2))\\
    &q_2 = -\dfrac{2\pi}{9 d}( ( \sqrt{3}\epsilon_{xx}\cos(\theta/2) + (6-3\epsilon_{yy})\sin(\theta/2), -3\epsilon_{yy}\cos(\theta/2) +(2\sqrt{3}- \sqrt{3}\epsilon_{xx})\sin(\theta/2))\\   
    &q_3 = -\dfrac{2\pi}{9 d}( ( \sqrt{3}\epsilon_{xx}\cos(\theta/2) - (6-3\epsilon_{xx})\sin(\theta/2), 3\epsilon_{yy}\cos(\theta/2) +(2\sqrt{3}- \sqrt{3}\epsilon_{xx})\sin(\theta/2)),
\end{align}

where we have consider that $\epsilon_{xy} =  \epsilon_{yx} = 0$. We adopt graphene-like parameters:  $\hbar v_f = 5.24$ meV$\AA$, $\beta = 1.57$, $\nu = 0.165$, $u = 79.7$ meV and $u' = 97$ meV, and $a = 2.46$ $\text{\AA}$ \cite{Koshino2018}. The Hamiltonian is diagonalized over the moiré Brillouin zone on a uniform grid of approximately 7500 k-points. The quantities calculated, such as Berry-curvature, Chern numbers, and electric polarization, and related functions, are computed using the modern theory of polarization \cite{Vanderbilt2018}, at half-filling and zero temperature.

\section*{Data availability}

The data that support the findings of this study are available from the corresponding author upon reasonable request.

\section*{Acknowledgements}

Funding is acknowledged from Fondecyt Regular 1230515 and Cedenna CIA250002. M.A.C. acknowledges Proyecto ANID Fondecyt Postdoctorado 3240112.

\section*{Competing interests}

The authors declare no competing interests.

\section*{Author contributions}

M. Castro performed the numerical calculations. A. S. Nunez conceived the project. F. Wolff, M. Castro, and A. S. Nunez participated in the elaboration of the figures. All authors participated in the discussion of the results and manuscript writing.


\begin{thebibliography}{10}
\urlstyle{rm}
\expandafter\ifx\csname url\endcsname\relax
  \def\url#1{\texttt{#1}}\fi
\expandafter\ifx\csname urlprefix\endcsname\relax\def\urlprefix{URL }\fi
\expandafter\ifx\csname doiprefix\endcsname\relax\def\doiprefix{DOI: }\fi
\providecommand{\bibinfo}[2]{#2}
\providecommand{\eprint}[2][]{\url{#2}}

\bibitem{Maekawa2017}
\bibinfo{editor}{Maekawa, S.}, \bibinfo{editor}{Valenzuela, S.~O.}, \bibinfo{editor}{Saitoh, E.} \& \bibinfo{editor}{Kimura, T.} (eds.) \emph{\bibinfo{title}{Spin Current}}.
\newblock Series on Semiconductor Science and Technology (\bibinfo{publisher}{Oxford University Press}, \bibinfo{address}{London, England}, \bibinfo{year}{2017}), \bibinfo{edition}{2} edn.

\bibitem{Xu2015}
\bibinfo{editor}{Xu, Y.}, \bibinfo{editor}{Awschalom, D.~D.} \& \bibinfo{editor}{Nitta, J.} (eds.) \emph{\bibinfo{title}{Handbook of Spintronics}}.
\newblock Handbook of Spintronics (\bibinfo{publisher}{Springer}, \bibinfo{address}{Dordrecht, Netherlands}, \bibinfo{year}{2015}), \bibinfo{edition}{1} edn.

\bibitem{Dey2022}
\bibinfo{author}{Dey, P.} \& \bibinfo{author}{Roy, J.~N.}
\newblock \emph{\bibinfo{title}{Spintronics}} (\bibinfo{publisher}{Springer}, \bibinfo{address}{Singapore, Singapore}, \bibinfo{year}{2022}), \bibinfo{edition}{2021} edn.

\bibitem{Nunez2014}
\bibinfo{author}{N{\'u}{\~n}ez, {\'A}.~S.}
\newblock \bibinfo{journal}{\bibinfo{title}{Theory of the piezo-spintronic effect}}.
\newblock {\emph{\JournalTitle{Solid State Communications}}} \textbf{\bibinfo{volume}{198}}, \bibinfo{pages}{18--21}, \doiprefix\url{10.1016/j.ssc.2013.10.018} (\bibinfo{year}{2014}).

\bibitem{Curie1880}
\bibinfo{author}{Curie, J.} \& \bibinfo{author}{Curie, P.}
\newblock \bibinfo{journal}{\bibinfo{title}{D{\'e}veloppement par compression de l'{\'e}lectricit{\'e} polaire dans les cristaux h{\'e}mi{\`e}dres {\`a} faces inclin{\'e}es}}.
\newblock {\emph{\JournalTitle{Bulletin de la Soci{\'e}t{\'e} min{\'e}ralogique de France}}} \textbf{\bibinfo{volume}{3}}, \bibinfo{pages}{90--93}, \doiprefix\url{10.3406/bulmi.1880.1564} (\bibinfo{year}{1880}).

\bibitem{Gautschi2002}
\bibinfo{author}{Gautschi, G.}
\newblock \emph{\bibinfo{title}{Piezoelectric Sensorics}} (\bibinfo{publisher}{Springer Berlin Heidelberg}, \bibinfo{year}{2002}).

\bibitem{Cady2018}
\bibinfo{author}{Cady, W.}
\newblock \emph{\bibinfo{title}{Piezoelectricity: Volume one: An introduction to the theory and applications of electromechanical phenomena in crystals}} (\bibinfo{publisher}{Dover Publications}, \bibinfo{address}{Mineola, NY}, \bibinfo{year}{2018}).

\bibitem{Ulloa2017}
\bibinfo{author}{Ulloa, C.}, \bibinfo{author}{Troncoso, R.~E.}, \bibinfo{author}{Bender, S.~A.}, \bibinfo{author}{Duine, R.~A.} \& \bibinfo{author}{Nunez, A.~S.}
\newblock \bibinfo{journal}{\bibinfo{title}{Piezospintronic effect in honeycomb antiferromagnets}}.
\newblock {\emph{\JournalTitle{Phys. Rev. B}}} \textbf{\bibinfo{volume}{96}}, \bibinfo{pages}{104419}, \doiprefix\url{10.1103/PhysRevB.96.104419} (\bibinfo{year}{2017}).

\bibitem{Castro2024}
\bibinfo{author}{Castro, M.}, \bibinfo{author}{Saéz, G.}, \bibinfo{author}{Vergara~Apaz, P.}, \bibinfo{author}{Allende, S.} \& \bibinfo{author}{Nunez, A.~S.}
\newblock \bibinfo{journal}{\bibinfo{title}{Toward fully multiferroic van der waals spinfets: Basic design and quantum calculations}}.
\newblock {\emph{\JournalTitle{Nano Letters}}} \textbf{\bibinfo{volume}{24}}, \bibinfo{pages}{7911–7918}, \doiprefix\url{10.1021/acs.nanolett.4c01146} (\bibinfo{year}{2024}).

\bibitem{Vergara2024}
\bibinfo{author}{Vergara, P.}, \bibinfo{author}{S{\'a}ez, G.}, \bibinfo{author}{Castro, M.}, \bibinfo{author}{Allende, S.} \& \bibinfo{author}{N{\'u}{\~n}ez, {\'A}.~S.}
\newblock \bibinfo{journal}{\bibinfo{title}{Emerging topological multiferroics from the 2d rice-mele model}}.
\newblock {\emph{\JournalTitle{npj 2D Materials and Applications}}} \textbf{\bibinfo{volume}{8}}, \doiprefix\url{10.1038/s41699-024-00478-5} (\bibinfo{year}{2024}).

\bibitem{Vanderbilt2018}
\bibinfo{author}{Vanderbilt, D.}
\newblock \emph{\bibinfo{title}{Berry phases in electronic structure theory}} (\bibinfo{publisher}{Cambridge University Press}, \bibinfo{address}{Cambridge, England}, \bibinfo{year}{2018}).

\bibitem{Avron1997}
\bibinfo{author}{Avron, J.~E.}, \bibinfo{author}{Berger, J.} \& \bibinfo{author}{Last, Y.}
\newblock \bibinfo{journal}{\bibinfo{title}{Piezoelectricity: Quantized charge transport driven by adiabatic deformations}}.
\newblock {\emph{\JournalTitle{Physical Review Letters}}} \textbf{\bibinfo{volume}{78}}, \bibinfo{pages}{511–514}, \doiprefix\url{10.1103/physrevlett.78.511} (\bibinfo{year}{1997}).

\bibitem{Saez2023}
\bibinfo{author}{Saez, G.}, \bibinfo{author}{Castro, M.~A.}, \bibinfo{author}{Allende, S.} \& \bibinfo{author}{Nunez, A.~S.}
\newblock \bibinfo{journal}{\bibinfo{title}{Model for nonrelativistic topological multiferroic matter}}.
\newblock {\emph{\JournalTitle{Physical Review Letters}}} \textbf{\bibinfo{volume}{131}}, \doiprefix\url{10.1103/physrevlett.131.226801} (\bibinfo{year}{2023}).

\bibitem{Saez2024}
\bibinfo{author}{S{\'a}ez, G.}, \bibinfo{author}{Vergara, P.}, \bibinfo{author}{Castro, M.}, \bibinfo{author}{Allende, S.} \& \bibinfo{author}{N{\'u}{\~n}ez, {\'A}.~S.}
\newblock \bibinfo{journal}{\bibinfo{title}{Ferrospintronic order in noncentrosymmetric antiferromagnets: An avenue toward spintronic-based computing, data storage, and energy harvesting}}.
\newblock {\emph{\JournalTitle{physica status solidi ({RRL}) Rapid Research Letters}}} \textbf{\bibinfo{volume}{19}}, \doiprefix\url{10.1002/pssr.202400292} (\bibinfo{year}{2024}).

\bibitem{Chen2025}
\bibinfo{author}{Chen, Y.}, \bibinfo{author}{Ji, J.}, \bibinfo{author}{Hong, L.}, \bibinfo{author}{Wan, X.} \& \bibinfo{author}{Xiang, H.}
\newblock \bibinfo{journal}{\bibinfo{title}{Generation of pure spin current with insulating antiferromagnetic materials}}.
\newblock {\emph{\JournalTitle{Physical Review Letters}}} \textbf{\bibinfo{volume}{135}}, \doiprefix\url{10.1103/n8d2-hjnd} (\bibinfo{year}{2025}).

\bibitem{Yao2024}
\bibinfo{author}{Yao, F.} \emph{et~al.}
\newblock \bibinfo{journal}{\bibinfo{title}{Moiré magnetism in crbr3 multilayers emerging from differential strain}}.
\newblock {\emph{\JournalTitle{Nature Communications}}} \textbf{\bibinfo{volume}{15}}, \doiprefix\url{10.1038/s41467-024-54870-2} (\bibinfo{year}{2024}).

\bibitem{Ciorciaro2023}
\bibinfo{author}{Ciorciaro, L.} \emph{et~al.}
\newblock \bibinfo{journal}{\bibinfo{title}{Kinetic magnetism in triangular moiré materials}}.
\newblock {\emph{\JournalTitle{Nature}}} \textbf{\bibinfo{volume}{623}}, \bibinfo{pages}{509–513}, \doiprefix\url{10.1038/s41586-023-06633-0} (\bibinfo{year}{2023}).

\bibitem{AkifKeskiner2024}
\bibinfo{author}{Akif~Keskiner, M.}, \bibinfo{author}{Ghaemi, P.}, \bibinfo{author}{Oktel, M.~O.} \& \bibinfo{author}{Erten, O.}
\newblock \bibinfo{journal}{\bibinfo{title}{Theory of moiré magnetism and multidomain spin textures in twisted mott insulator–semimetal heterobilayers}}.
\newblock {\emph{\JournalTitle{Nano Letters}}} \textbf{\bibinfo{volume}{24}}, \bibinfo{pages}{8575–8579}, \doiprefix\url{10.1021/acs.nanolett.4c01574} (\bibinfo{year}{2024}).

\bibitem{Liu2024}
\bibinfo{author}{Liu, J.}, \bibinfo{author}{Zhang, X.} \& \bibinfo{author}{Lu, G.}
\newblock \bibinfo{journal}{\bibinfo{title}{Moiré magnetism and moiré excitons in twisted crsbr bilayers}}.
\newblock {\emph{\JournalTitle{Proceedings of the National Academy of Sciences}}} \textbf{\bibinfo{volume}{122}}, \doiprefix\url{10.1073/pnas.2413326121} (\bibinfo{year}{2024}).

\bibitem{Slot2023}
\bibinfo{author}{Slot, M.~R.} \emph{et~al.}
\newblock \bibinfo{journal}{\bibinfo{title}{A quantum ruler for orbital magnetism in moiré quantum matter}}.
\newblock {\emph{\JournalTitle{Science}}} \textbf{\bibinfo{volume}{382}}, \bibinfo{pages}{81–87}, \doiprefix\url{10.1126/science.adf2040} (\bibinfo{year}{2023}).

\bibitem{Fumega2023}
\bibinfo{author}{Fumega, A.~O.} \& \bibinfo{author}{Lado, J.~L.}
\newblock \bibinfo{journal}{\bibinfo{title}{Moiré-driven multiferroic order in twisted crcl3, crbr3 and cri3 bilayers}}.
\newblock {\emph{\JournalTitle{2D Materials}}} \textbf{\bibinfo{volume}{10}}, \bibinfo{pages}{025026}, \doiprefix\url{10.1088/2053-1583/acc671} (\bibinfo{year}{2023}).

\bibitem{Abouelkomsan2024}
\bibinfo{author}{Abouelkomsan, A.}, \bibinfo{author}{Bergholtz, E.~J.} \& \bibinfo{author}{Chatterjee, S.}
\newblock \bibinfo{journal}{\bibinfo{title}{Multiferroicity and topology in twisted transition metal dichalcogenides}}.
\newblock {\emph{\JournalTitle{Physical Review Letters}}} \textbf{\bibinfo{volume}{133}}, \doiprefix\url{10.1103/physrevlett.133.026801} (\bibinfo{year}{2024}).

\bibitem{Sun2022}
\bibinfo{author}{Sun, W.} \emph{et~al.}
\newblock \bibinfo{journal}{\bibinfo{title}{Labr2 bilayer multiferroic moiré superlattice with robust magnetoelectric coupling and magnetic bimerons}}.
\newblock {\emph{\JournalTitle{npj Computational Materials}}} \textbf{\bibinfo{volume}{8}}, \doiprefix\url{10.1038/s41524-022-00833-4} (\bibinfo{year}{2022}).

\bibitem{Sun2023}
\bibinfo{author}{Sun, W.} \emph{et~al.}
\newblock \bibinfo{journal}{\bibinfo{title}{Quantized movement of magnetic skyrmions in moiré multiferroic heterostructures}}.
\newblock {\emph{\JournalTitle{Physical Review B}}} \textbf{\bibinfo{volume}{107}}, \doiprefix\url{10.1103/physrevb.107.184439} (\bibinfo{year}{2023}).

\bibitem{Bistritzer2011}
\bibinfo{author}{Bistritzer, R.} \& \bibinfo{author}{MacDonald, A.~H.}
\newblock \bibinfo{journal}{\bibinfo{title}{Moiré bands in twisted double-layer graphene}}.
\newblock {\emph{\JournalTitle{Proceedings of the National Academy of Sciences}}} \textbf{\bibinfo{volume}{108}}, \bibinfo{pages}{12233–12237}, \doiprefix\url{10.1073/pnas.1108174108} (\bibinfo{year}{2011}).

\bibitem{Andrei2020}
\bibinfo{author}{Andrei, E.~Y.} \& \bibinfo{author}{MacDonald, A.~H.}
\newblock \bibinfo{journal}{\bibinfo{title}{Graphene bilayers with a twist}}.
\newblock {\emph{\JournalTitle{Nature Materials}}} \textbf{\bibinfo{volume}{19}}, \bibinfo{pages}{1265–1275}, \doiprefix\url{10.1038/s41563-020-00840-0} (\bibinfo{year}{2020}).

\bibitem{LopesdosSantos2012}
\bibinfo{author}{Lopes~dos Santos, J. M.~B.}, \bibinfo{author}{Peres, N. M.~R.} \& \bibinfo{author}{Castro~Neto, A.~H.}
\newblock \bibinfo{journal}{\bibinfo{title}{Continuum model of the twisted graphene bilayer}}.
\newblock {\emph{\JournalTitle{Physical Review B}}} \textbf{\bibinfo{volume}{86}}, \doiprefix\url{10.1103/physrevb.86.155449} (\bibinfo{year}{2012}).

\bibitem{LopesdosSantos2007}
\bibinfo{author}{Lopes dos Santos, J. M.~B.}, \bibinfo{author}{Peres, N. M.~R.} \& \bibinfo{author}{Castro Neto, A.~H.}
\newblock \bibinfo{journal}{\bibinfo{title}{Graphene bilayer with a twist: Electronic structure}}.
\newblock {\emph{\JournalTitle{Physical Review Letters}}} \textbf{\bibinfo{volume}{99}}, \doiprefix\url{10.1103/physrevlett.99.256802} (\bibinfo{year}{2007}).

\bibitem{He2013}
\bibinfo{author}{He, W.-Y.}, \bibinfo{author}{Chu, Z.-D.} \& \bibinfo{author}{He, L.}
\newblock \bibinfo{journal}{\bibinfo{title}{Chiral tunneling in a twisted graphene bilayer}}.
\newblock {\emph{\JournalTitle{Physical Review Letters}}} \textbf{\bibinfo{volume}{111}}, \doiprefix\url{10.1103/physrevlett.111.066803} (\bibinfo{year}{2013}).

\bibitem{Watson2023}
\bibinfo{author}{Watson, A.~B.}, \bibinfo{author}{Kong, T.}, \bibinfo{author}{MacDonald, A.~H.} \& \bibinfo{author}{Luskin, M.}
\newblock \bibinfo{journal}{\bibinfo{title}{Bistritzer–macdonald dynamics in twisted bilayer graphene}}.
\newblock {\emph{\JournalTitle{Journal of Mathematical Physics}}} \textbf{\bibinfo{volume}{64}}, \doiprefix\url{10.1063/5.0115771} (\bibinfo{year}{2023}).

\bibitem{Becker2024}
\bibinfo{author}{Becker, S.} \& \bibinfo{author}{Zworski, M.}
\newblock \bibinfo{journal}{\bibinfo{title}{From the chiral model of tbg to the bistritzer–macdonald model}}.
\newblock {\emph{\JournalTitle{Journal of Mathematical Physics}}} \textbf{\bibinfo{volume}{65}}, \doiprefix\url{10.1063/5.0174062} (\bibinfo{year}{2024}).

\bibitem{Serlin2020}
\bibinfo{author}{Serlin, M.} \emph{et~al.}
\newblock \bibinfo{journal}{\bibinfo{title}{Intrinsic quantized anomalous hall effect in a moiré heterostructure}}.
\newblock {\emph{\JournalTitle{Science}}} \textbf{\bibinfo{volume}{367}}, \bibinfo{pages}{900–903}, \doiprefix\url{10.1126/science.aay5533} (\bibinfo{year}{2020}).

\bibitem{Xiao2020}
\bibinfo{author}{Xiao, Y.}, \bibinfo{author}{Liu, J.} \& \bibinfo{author}{Fu, L.}
\newblock \bibinfo{journal}{\bibinfo{title}{Moiré is more: Access to new properties of two-dimensional layered materials}}.
\newblock {\emph{\JournalTitle{Matter}}} \textbf{\bibinfo{volume}{3}}, \bibinfo{pages}{1142–1161}, \doiprefix\url{10.1016/j.matt.2020.07.001} (\bibinfo{year}{2020}).

\bibitem{He2020}
\bibinfo{author}{He, W.-Y.}, \bibinfo{author}{Goldhaber-Gordon, D.} \& \bibinfo{author}{Law, K.~T.}
\newblock \bibinfo{journal}{\bibinfo{title}{Giant orbital magnetoelectric effect and current-induced magnetization switching in twisted bilayer graphene}}.
\newblock {\emph{\JournalTitle{Nature Communications}}} \textbf{\bibinfo{volume}{11}}, \doiprefix\url{10.1038/s41467-020-15473-9} (\bibinfo{year}{2020}).

\bibitem{Zhang2022}
\bibinfo{author}{Zhang, C.-P.} \emph{et~al.}
\newblock \bibinfo{journal}{\bibinfo{title}{Giant nonlinear hall effect in strained twisted bilayer graphene}}.
\newblock {\emph{\JournalTitle{Physical Review B}}} \textbf{\bibinfo{volume}{106}}, \doiprefix\url{10.1103/physrevb.106.l041111} (\bibinfo{year}{2022}).

\bibitem{Arora2021}
\bibinfo{author}{Arora, A.}, \bibinfo{author}{Kong, J.~F.} \& \bibinfo{author}{Song, J. C.~W.}
\newblock \bibinfo{journal}{\bibinfo{title}{Strain-induced large injection current in twisted bilayer graphene}}.
\newblock {\emph{\JournalTitle{Physical Review B}}} \textbf{\bibinfo{volume}{104}}, \doiprefix\url{10.1103/physrevb.104.l241404} (\bibinfo{year}{2021}).

\bibitem{Cao2020}
\bibinfo{author}{Cao, J.}, \bibinfo{author}{Qi, F.}, \bibinfo{author}{Yang, H.} \& \bibinfo{author}{Jin, G.}
\newblock \bibinfo{journal}{\bibinfo{title}{Monolayer-gap modulated topological phases in twisted bilayer graphene}}.
\newblock {\emph{\JournalTitle{Physical Review B}}} \textbf{\bibinfo{volume}{101}}, \doiprefix\url{10.1103/physrevb.101.155419} (\bibinfo{year}{2020}).

\bibitem{Yu2020}
\bibinfo{author}{Yu, J.} \& \bibinfo{author}{Liu, C.-X.}
\newblock \bibinfo{journal}{\bibinfo{title}{Piezoelectricity and topological quantum phase transitions in two-dimensional spin-orbit coupled crystals with time-reversal symmetry}}.
\newblock {\emph{\JournalTitle{Nature Communications}}} \textbf{\bibinfo{volume}{11}}, \doiprefix\url{10.1038/s41467-020-16058-2} (\bibinfo{year}{2020}).

\bibitem{KingSmith1993}
\bibinfo{author}{King-Smith, R.~D.} \& \bibinfo{author}{Vanderbilt, D.}
\newblock \bibinfo{journal}{\bibinfo{title}{Theory of polarization of crystalline solids}}.
\newblock {\emph{\JournalTitle{Physical Review B}}} \textbf{\bibinfo{volume}{47}}, \bibinfo{pages}{1651–1654}, \doiprefix\url{10.1103/physrevb.47.1651} (\bibinfo{year}{1993}).

\bibitem{Peng2022}
\bibinfo{author}{Peng, R.} \& \bibinfo{author}{Liu, J.}
\newblock \bibinfo{journal}{\bibinfo{title}{Topological piezoelectric response in moiré graphene systems}}.
\newblock {\emph{\JournalTitle{Physical Review Research}}} \textbf{\bibinfo{volume}{4}}, \doiprefix\url{10.1103/physrevresearch.4.l032006} (\bibinfo{year}{2022}).

\bibitem{Jadaun2023}
\bibinfo{author}{Jadaun, P.} \& \bibinfo{author}{Soreé, B.}
\newblock \bibinfo{journal}{\bibinfo{title}{Review of orbital magnetism in graphene-based moiré materials}}.
\newblock {\emph{\JournalTitle{Magnetism}}} \textbf{\bibinfo{volume}{3}}, \bibinfo{pages}{245–258}, \doiprefix\url{10.3390/magnetism3030019} (\bibinfo{year}{2023}).

\bibitem{Sharpe2019}
\bibinfo{author}{Sharpe, A.~L.} \emph{et~al.}
\newblock \bibinfo{journal}{\bibinfo{title}{Emergent ferromagnetism near three-quarters filling in twisted bilayer graphene}}.
\newblock {\emph{\JournalTitle{Science}}} \textbf{\bibinfo{volume}{365}}, \bibinfo{pages}{605–608}, \doiprefix\url{10.1126/science.aaw3780} (\bibinfo{year}{2019}).

\bibitem{Tschirhart2021}
\bibinfo{author}{Tschirhart, C.~L.} \emph{et~al.}
\newblock \bibinfo{journal}{\bibinfo{title}{Imaging orbital ferromagnetism in a moiré chern insulator}}.
\newblock {\emph{\JournalTitle{Science}}} \textbf{\bibinfo{volume}{372}}, \bibinfo{pages}{1323–1327}, \doiprefix\url{10.1126/science.abd3190} (\bibinfo{year}{2021}).

\bibitem{Guerci2021}
\bibinfo{author}{Guerci, D.}, \bibinfo{author}{Simon, P.} \& \bibinfo{author}{Mora, C.}
\newblock \bibinfo{journal}{\bibinfo{title}{Moiré lattice effects on the orbital magnetic response of twisted bilayer graphene and condon instability}}.
\newblock {\emph{\JournalTitle{Physical Review B}}} \textbf{\bibinfo{volume}{103}}, \doiprefix\url{10.1103/physrevb.103.224436} (\bibinfo{year}{2021}).

\bibitem{Chen2023}
\bibinfo{author}{Chen, J.}, \bibinfo{author}{Liu, C.} \& \bibinfo{author}{Li, R.}
\newblock \bibinfo{journal}{\bibinfo{title}{Valley-selective polarization in twisted bilayer graphene controlled by a counter-rotating bicircular laser field}}.
\newblock {\emph{\JournalTitle{Photonics}}} \textbf{\bibinfo{volume}{10}}, \bibinfo{pages}{516}, \doiprefix\url{10.3390/photonics10050516} (\bibinfo{year}{2023}).

\bibitem{Farajollahpour2017}
\bibinfo{author}{Farajollahpour, T.} \& \bibinfo{author}{Phirouznia, A.}
\newblock \bibinfo{journal}{\bibinfo{title}{The role of the strain induced population imbalance in valley polarization of graphene: Berry curvature perspective}}.
\newblock {\emph{\JournalTitle{Scientific Reports}}} \textbf{\bibinfo{volume}{7}}, \doiprefix\url{10.1038/s41598-017-18238-5} (\bibinfo{year}{2017}).

\bibitem{Jing2023}
\bibinfo{author}{Jing, Y.}, \bibinfo{author}{Jingxue, D.}, \bibinfo{author}{Weijun, F.} \& \bibinfo{author}{Lijie, S.}
\newblock \bibinfo{journal}{\bibinfo{title}{Vertical electric-field controlled electronic properties of mose2/mnpse3 van der waals heterojunction}}.
\newblock {\emph{\JournalTitle{Physica B: Condensed Matter}}} \textbf{\bibinfo{volume}{665}}, \bibinfo{pages}{415076}, \doiprefix\url{10.1016/j.physb.2023.415076} (\bibinfo{year}{2023}).

\bibitem{Li2013}
\bibinfo{author}{Li, X.}, \bibinfo{author}{Cao, T.}, \bibinfo{author}{Niu, Q.}, \bibinfo{author}{Shi, J.} \& \bibinfo{author}{Feng, J.}
\newblock \bibinfo{journal}{\bibinfo{title}{Coupling the valley degree of freedom to antiferromagnetic order}}.
\newblock {\emph{\JournalTitle{Proceedings of the National Academy of Sciences}}} \textbf{\bibinfo{volume}{110}}, \bibinfo{pages}{3738–3742}, \doiprefix\url{10.1073/pnas.1219420110} (\bibinfo{year}{2013}).

\bibitem{Rahman2021}
\bibinfo{author}{Rahman, S.}, \bibinfo{author}{Torres, J.~F.}, \bibinfo{author}{Khan, A.~R.} \& \bibinfo{author}{Lu, Y.}
\newblock \bibinfo{journal}{\bibinfo{title}{Recent developments in van der waals antiferromagnetic 2d materials: Synthesis, characterization, and device implementation}}.
\newblock {\emph{\JournalTitle{ACS Nano}}} \textbf{\bibinfo{volume}{15}}, \bibinfo{pages}{17175–17213}, \doiprefix\url{10.1021/acsnano.1c06864} (\bibinfo{year}{2021}).

\bibitem{Calder2021}
\bibinfo{author}{Calder, S.}, \bibinfo{author}{Haglund, A.~V.}, \bibinfo{author}{Kolesnikov, A.~I.} \& \bibinfo{author}{Mandrus, D.}
\newblock \bibinfo{journal}{\bibinfo{title}{Magnetic exchange interactions in the van der waals layered antiferromagnet {MnPSe}$_{3}$}}.
\newblock {\emph{\JournalTitle{Physical Review B}}} \textbf{\bibinfo{volume}{103}}, \doiprefix\url{10.1103/physrevb.103.024414} (\bibinfo{year}{2021}).

\bibitem{Lee2020}
\bibinfo{author}{Lee, K.~W.} \& \bibinfo{author}{Lee, C.~E.}
\newblock \bibinfo{journal}{\bibinfo{title}{Quantum valley hall effect in wide-gap semiconductor sic monolayer}}.
\newblock {\emph{\JournalTitle{Scientific Reports}}} \textbf{\bibinfo{volume}{10}}, \doiprefix\url{10.1038/s41598-020-61906-2} (\bibinfo{year}{2020}).

\bibitem{Koshino2018}
\bibinfo{author}{Koshino, M.} \emph{et~al.}
\newblock \bibinfo{journal}{\bibinfo{title}{Maximally localized wannier orbitals and the extended hubbard model for twisted bilayer graphene}}.
\newblock {\emph{\JournalTitle{Physical Review X}}} \textbf{\bibinfo{volume}{8}}, \doiprefix\url{10.1103/physrevx.8.031087} (\bibinfo{year}{2018}).

\end{thebibliography}
\end{document}